Discussion on the nucleon decay experiments

S.Miyake
Osaka City University, Osaka, Japan

Abstract

The results of KGF experiments are compared with the results of Soudan 2 and Frejus experiments and reasonable agreements are obtained. Nevertheless, the results of Super-Kamiokande (SK) are somewhat different from the other experiments and some points in SK are discussed.

Recently, there are several reports of nucleon decay experiments. [Ref.1-9] Therefore, it is important to discuss what has been known already and what should be done for future. However, the typical conclusion of these papers is "No evidence for a nucleon decay signal is observed and we report lifetime lower limit at 90 % C.L.", except our K.G.F. report. [1] Then here, along with the point of issue written in our previous paper " The apparent contradiction between these conclusions does not mean a complete disagreement between the observations, but a clue to this question comes from the interpretation of the experimental results", I would like to try to get tentative conclusion by the comparison of these papers.

There are two types of observations, namely, one is the direct observation of the events by the use of calorimeter type detectors and the other is the indirect observation through the observation of Cerenkov radiation in water. It may be better to discuss first about the direct observations to make the image of the observed nucleon decay event clear. Then the discussion for the results of indirect observations will be done.

I-1) Comparison of the result of Soudan 2 [5,9] with that of K,G.F. [1]

For the sake of the discussion in a same standard, as the first approximation, we take the same ratio of the fiducial exposure factor to the total exposure for both the experiments. Then, the effective exposure factors are 3.3 kty for Soudan 2 and 1.7 kty for KGF respectively.

Soudan 2 has shown one beautiful picture of the candidate event p - e+ + K0s, which is very similar to that of KGF. According to the paper, the detection efficiency including the branching ratio is 15 %. Hence, the rate of occurrence of such candidate events is one per 0.50 kty in Soudan 2. Similar calculation in KGF 3 events in 1.7 kty with nearly full efficiency, gives the rate one per 0.56 kty. Although the statistical errors in both experiments are large, they show very good agreement to each other.

Nevertheless, there are large difference in their estimated backgrounds which is 0.7 in Soudan 2 and 0.0l in KGF respectively. The large difference between them are mainly due to two reasons, namely, one is the estimated Fermi momentum which is used in the Monte Carlo simulation to calculate background and the other is the accuracy of observation by which the level of background is estimated. Actually, the upper limit of the Fermi momentum used in Soudan 2 is as large as 500 MeV/c and it is practically less than 200 MeV/c in KGF. This large difference makes about 10 times of the difference in the background level. On the other hand, the accuracy in energy measurement of observed tracks is better in KGF than in Soudan 2, though the impression from the sizes of their counters is opposite. The square shape instead of honeycomb is favor for track length method for energy measurement. It makes another factor of about 3 times in the difference of their background levels. Thus, combined above two factors, one may understand rough idea about the large difference in the background levels of both the experiments.

Now, it is clear that the problem of background for the nucleon decay event is largely depends on the value of Fermi momentum. If Fermi momentum is surely lower than 200 MeV/c, the event observed in Soudan 2 is highly probable to be a proton decay event because of the low background. At the same time, it also supports the result of KGF.

I-2) Estimation of Fermi momentum in KGF Experiment.

In the previous KGF paper, we have used only the experimental facts verified in our experiment and others to estimate the background level to the candidate events, because Monte

Carlo simulation has large ambiguities to estimate the shape of the proton decay event and also the mimic event from atmospheric neutrino interaction. As the main origin of the above ambiguity is the Fermi motion of the nucleons in iron nucleus, the value of the Fermi momentum is estimated from the following three experimental facts observed in KGF experiment.

(i) The energy of kaons observed as the candidate events for the decay mode p - anti-nu + K+ are less than 700 MeV, and most of them are very low energy, not even coming out from the iron plates of the detectors.

(ii) The candidate events having two visible decay particles show back to back configuration which makes clearly separated group of nucleon decay candidate events from atmospheric neutrino events.

(iii) Four pions observed as candidates of p - anti-nu + pi+ decay, supposed to be emitted to back hemisphere from the running proton, are distributed within about 30 MeV width.

If the Fermi momentum, p, could be approximated by the following expression,

$$F(p) \, dp = A \, p^2 \exp(-p^2/p_0^2) \, dp \qquad (1)$$

the value of $p_0$ may be around 100 MeV/c to explain above three experimental results.

As mentioned already in the previous section, the size of the upper limit of the Fermi motion of nucleons, around 200 MeV/c or 500 MeV/c, makes essential role in deforming the shape of nucleon decay event and also the sudden increase of the probability of mimic event from atmospheric neutrino interaction. The effect is due to the value of Fermi momentum whether it is very small compared with the momentum of nucleon decay products or comparable with them. It may be necessary to state a clear conclusion of KGF experiment again that Fermi momentum is as small as less than 200MeV/c, and the background for the accurate candidate event is the order of a few percents.

I-3) General view of the comparison

KGF have observed many candidate events in the exposure factor of 1.7 kty as shown in the previous paper. [1] Because of the great depth of the operational site, total number of the observed events is only about 10 per day and all of them including atmospheric muons are analyzed. In case of shallow depth, the total number of events is as large as about $10^5$ times more, and naturally, one has to use a set of strict selection criteria to obtain candidate events. The efficiency of such selection in Soudan 2 is as small as 10 - 15 %. Therefore, even having the exposure factor 3.3 kty, the effective exposure factor of Soudan 2 is about 0.3 - 0.4 kty and difficult to get comparable number of events to KGF. The results of partial lifetime for the other various decay modes obtained in KGF experiment also do not contradict to the lower limits obtained by Soudan 2.

II-1) Comparison of the result of Frejus experiment [2,3,4] with KGF [1]

In their paper "Results for nucleon decay modes into anti-neutrino + meson" detection efficiencies are 10 % - 16 % for various decay modes and the exposure factor is 1.3 kty. Therefore, the effective exposure factor is only about 0.1 to 0.2 kty. It is too small to compare with the observation of 1.7 kty of KGF. The paper shows one clear picture of K-mu decay as a candidate of p - anti-nu + K+. The rate of occurrence is one event per 0.1 kty which agrees well to that of KGF. The result of KGF is, 10 events including 2 background events in 1.7 kty with efficiency x branching ratio, 0.6, namely, one event per 0.12 kty. In case of Frejus, one event is given as background (100 %), but it can be reduced to less than 0.5 easily by reducing the width of energy bin.

II-2) On the decay modes with charged leptons

The decay modes give total view, e+ or mu+ and pi or K or other mesons, so, we can check kinematics of each events, namely, candidate events should have the fixed conditions. There is some difference in the conditions. In KGF they are SUM(P) = Pf, and SUM(E) = Et = Mn-B, where Pf is Fermi momentum which is less than about 200 MeV/c, and B is a binding energy, 8 MeV. Allowance for the candidate events in Et is 10 - 15% depend upon the accuracy of the event. However, in case of Frejus, the former evaluation is same but the value of Pf seems to be allowed up to 500 MeV/c. The latter equation is different and written as [SUM(E)]^2 = mn^2 + pf^2, where mn is the nucleon mass inside the nuclear potential. Although the details are not written, probably, their expression may be similar to our simple expression.

In the table which shows lower limit of nucleon lifetime, the figures are understood as (i) the effective exposure factor is very small, for example, efficiency 0.2 x exposure factor (written as sensitivity) 1.6 kty = 0.3 kty, (ii) the background is comparable to the number of candidate events. Since the details of individual candidate events are not shown, further discussion is not possible. However, their results do not affect to the result of KGF which is about 5 times more in the effective exposure factor.

III-1) On the results of water Cerenkov detectors [6,7,8]

About the results of Kamiokande (K) and IMB are discussed already in our previous paper, in particular the results of Kamiokande agrees with our KGF result very well and used to verify the existence of proton decay in decay modes, p - anti-nu + K and p - anti-nu + pi together with KGF data. However, recent reports from Super-Kamiokande (SK) looks somewhat different, as is expressed as "no evidence for proton decay was observed". They may be showing some contradiction to the results of KGF experiment and others. Therefore, in this section, two items will be discussed, (i) problems in SK observation and (ii) Possibility of the results of SK to agree with the result of other experiment.

III-2) Some details of water Cerenkov detectors

Although it is difficult to estimate the character of the other experiment, assuming spherical shape of PMs of 50 cm diameter, there may be following conditions in SK observation.

(i) PMs in water may not be so stable. After some years run, there may be some numbers of PMs gone to out of order. If there is no repairing system, they may become some source of errors in the observation.

(ii) The definition of fiducial volume, 1.8 m inside from the top of PMs may be too short, particularly for gamma rays. There must be enough space for pair creation (9/7 r.l.), development of e.m. shower (several r.l.) and spread of Cerenkov light. If the border is taken another 2 m inside, the fiducial volume will be about 60 % of the paper. Therefore, two methods with about 5 % detection efficiency in SK have the effective exposure factor of only about 0.6 kty. even if there is no candidate event, it may be a problem of small number statistics.

(iii) The ratio of photo sensitive area at the PM plane will change from 40 % for vertical light to nearly 100 % for very inclined light which is a very complicated function of stereoscopic two angles. Not only the effective area of PM to the light but shadow of other PM on the PM also has to be taken into account for individual PM tubes. Moreover inner surface of glass of PM will give a strong reflection. For example, for the vertical light to PM plane, more than half of the light falling to the peripheral part of the tube will get total reflection. Even at the weakest reflection near the top of the tube, thin metal coating will act as half mirror. In between the above, the light get inside PM with deflection will hit again other part of photo sensitive layer. Thus, photo sensitivity is not equal by the position of the PM. After get main pulse, there is a delayed pulse of 1 - 2 ns by reflection of surrounding PMs. Rough estimation of the delayed pulse to the main pulse are 5 - 10 % in K and 20 - 30 % in SK. The reflection effect is naturally different by the position of PM at plane part and at near corner. The absorption of light at photosensitive layer is abbreviated.

(iv) SK papers have given four methods of calibration for the estimation of energy of the event observed by Cerenkov light. However, none of them looks for me to give the accuracy of 2.5 % as written in the paper, because of the following reasons. (a) Linac electron beam; it may be used at the beginning of the experiment under the limited condition but it cannot be a calibration for various cases in a long run. (b) dE/dx of slow muon will be disturbed by scattering of the muon. (c) The energy distribution of the decay electron from stopping muon is dull and low energy with large effect of scattering. The statistical comparison does not mean a correct calibration for each event, (d) The two light rings from pi0 overlap on each other and they will be difficult to separate into correct two rings. As mentioned in the previous section, it may needs a very complicated system of various types of calibrations due to the position of vertex and the direction of the event.

(v) The accuracy of the position of vertex, 17 cm, in SK is also unlikely because of the fluctuations in the development of e.m. cascade showers, size of the PM tubes, disturbance by the delayed pulses due to reflected light and deflected light as mentioned in (iii).

Ill-3) In general, the direct observation is always superior to the indirect observation. In case of direct observation, the figures of the event together with ionization and timing of every counter shown in the paper inform all the data. They are simple and clear, therefore, the readers of the paper are possible to discuss about the event at the same level of authors. Nevertheless, in case of water Cerenkov detector, there are many complicated processes to reconstruct the event under various assumptions. As mentioned already, the calibration and analysis of the water Cerenkov is not easy, however, the first check for the experiment may be to find out the similar events which have been observed already in the direct observations. If suppose one experiment with 10 to 20 % error in the energy region of a few hundred MeV/c, the peaks in the momentum distribution of muons from K+ decay and pions from P - anti-nu + pi+ will not be found. The systematic deviation of calibration of energy by about 10% may let the candidate events for P - e+ + pi0 escape.

IV   Conclusion

The comparison of KGF results with the other experiments show reasonable agreements except in case of SK. Although it is necessary to discuss still upon the points discussed about SK in III, tentative conclusion at the moment, is that there is no needs to change the conclusion of KGF experiment that protons are decaying with the lifetime of about $1 \times 10^{31}$ years through wide spectrum of decay mode.

V   Acknowledgement

The author is grateful to the member of KGF group and Prof. M.G.K.Menon who organized and encouraged India-Japan Collaboration for 40 years.